\documentclass[prl,twocolumn,aps,superscriptaddress,showpacs,floatfix]{revtex4}
\usepackage{amsmath}
\usepackage{graphicx}
\newcommand{\bis}{ Bi$_2$Sr$_2$CaCu$_2$O$_{8+\delta}\;$}

\newcommand{\sgn}{{\rm sgn}}

\begin{document}
\title{Sculpting microscopic magnetic flux landscapes in a Bi$_2$Sr$_2$CaCu$_2$O$_{8+\delta}$ vortex lens}

\author{D. Cole}
\affiliation{Department of Physics, University of Bath, Claverton
Down, Bath, BA2 7AY, UK}

\author{S.J. Bending}
\affiliation{Department of Physics, University of Bath, Claverton
Down, Bath, BA2 7AY, UK}

\author{Sergey Savel'ev}
\affiliation{Frontier Research System, The Institute of Physical
and Chemical Research (RIKEN), Wako-shi, Saitama 351-0198, Japan}
\affiliation{Department of Physics, Loughborough University, Loughborough LE11 3TU, UK}

\author{T. Tamegai}
\affiliation{Department of Applied Physics, The University of
Tokyo, Hongo, Bunkyo-ku, Tokyo 113-8656, Japan}

\author{Franco Nori}
\affiliation{Frontier Research System, The Institute of Physical
and Chemical Research (RIKEN), Wako-shi, Saitama, 351-0198, Japan}
\affiliation{Center for Theoretical Physics, Department of Physics, University\\
of Michigan, Ann Arbor, MI, 48109-1040,USA}

\date{\today}

\begin{abstract}
We demonstrate experimentally that the micromagnetic profile of
the out-of-plane component of magnetic induction, $B_z$, in the
crossing lattices regime of layered superconductors can be
manipulated by varying the in-plane magnetic field,
$H_{\parallel}$.  Moving Josephson vortices drag/push pancake vortex stacks, and the magnetic profile, $B_{z}(x)$, 
can be controllably sculpted across the entire single crystal sample. Depending on the $H$-history
and temperature we can increase or decrease the flux density at
the center and near the edges of the crystal by as
much as $\sim$ 40\%, realising both ``convex'' and ``concave'' magnetic
flux lenses. Our experimental results are well described by
molecular dynamics simulations.

\end{abstract}

\vspace{3cm}  \maketitle

In recent years dramatic progress has been made in the control of
{\it static} flux structures in superconductors by the
introduction of artificial vortex pinning sites (e.g. antidots and
ferromagnetic dots)
\cite{baert95,martin97,morgan98,bael99,lyuksyutov98}.  The next major challenge is
to achieve {\it dynamic} vortex control, so that different flux profiles
can be realised in the same superconducting sample.
Control of vortex motion has recently been proposed \cite{vortex} and demonstrated
\cite{kwok-2003,ourscience,moshchalkov,togawa,roger}
in so-called 'ratchet devices' which incorporate a spatially-asymmetric 'ratchet potential'
to achieve rectification of ac drives.
However, a spatially asymmetric ratchet substrate
is not a fundamental requirement for ratchet operation, and recent proposals
\cite{nature_materials,prl} have described
novel methods to control the motion of tiny particles in a binary mixture by
the dragging of one component by the other.
Here we describe the first such experimental implementation of a flux lens in the binary
vortex system present in highly anisotropic layered superconductors under tilted magnetic fields.

Direct visualization
\cite{tonomura,bending,vlasko,tamegai,bolle91} has revealed that a
tilted (away from the crystalline $c$-axis) magnetic field
penetrates the highly anisotropic \bis (BSCCO) superconductor in two
interpenetrating vortex arrays, known as 'crossing' vortex
lattices \cite{bul,kosh}.  One vortex sublattice consists of
stacks of pancake vortices (PVs) aligned along the $c$-axis, while
the other sublattice is formed by Josephson vortices (JVs)
confined between CuO$_2$ layers.  Superconducting currents
generated by JVs deform stacks of PVs, resulting in a mutual
attraction between PVs and JVs \cite{kosh,sav}.
This has been experimentally confirmed
\cite{tonomura,bending,vlasko,tamegai,bolle91} by the observation of PV chains
which 'decorate' underlying JV stacks in tilted magnetic fields.
JVs are usually very weakly pinned and can easily be driven by
changing the in-plane magnetic field $H_{{\parallel}}$ ($\equiv H_{ab}$), dragging
PVs along with them \cite{icl}. This joint JV-PV motion can be used to
develop \cite{nature_materials} vortex-pumps, vortex-diodes and
vortex-lenses which have two clear advantages over other existing
ratchet devices. (i) The motion of vortices can be controlled
without the need for nanofabricated samples with {\it fixed
spatial asymmetry}, and (ii) the focusing efficiency can be easily
varied by changing either the PV density (via the corresponding
magnetic field component $H_z$) and/or temperature.  Here we
describe how to implement such lenses experimentally.

\begin{figure}
  \includegraphics[width=1 \linewidth]{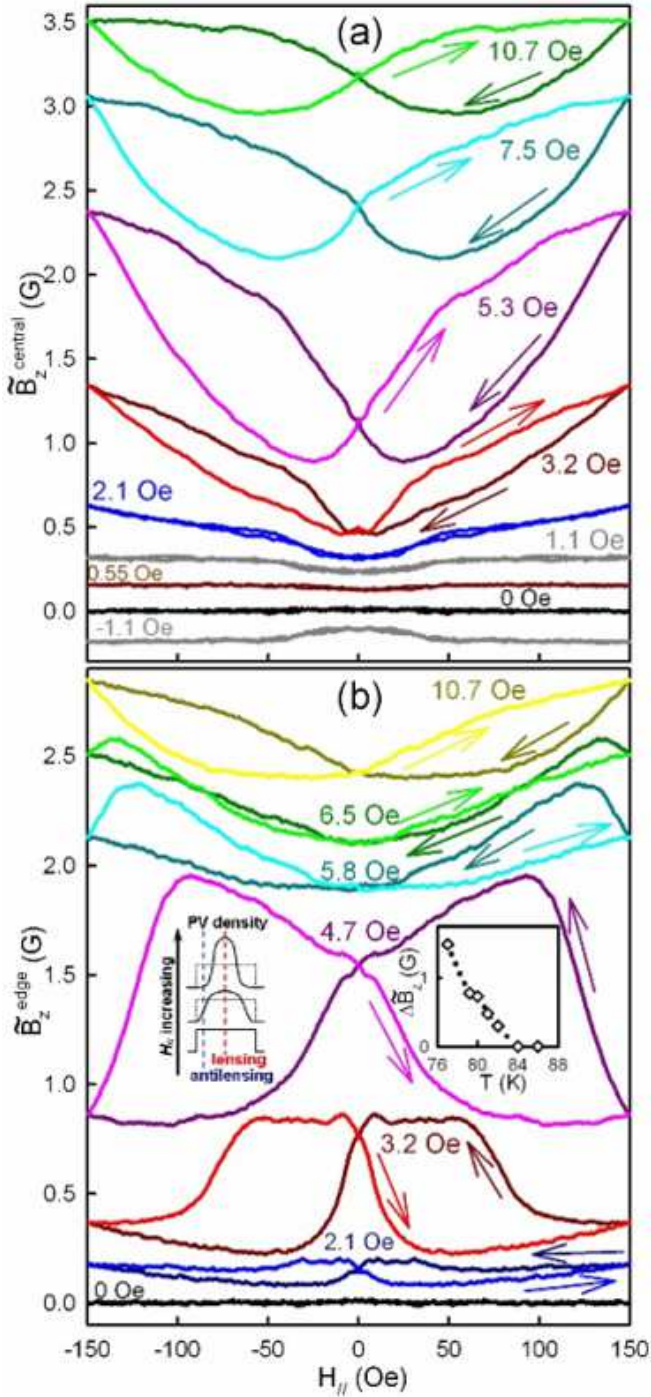}
  \caption{(Color online) Changes in the local out-of-plane magnetic induction, $\widetilde{B}_z$ (vertically offset for clarity), versus the in-plane magnetic field, $H_{{\parallel}}$, measured near the center (a) and near the edge (b) of the sample at 77 K. Arrows indicate the direction of the $H_{{\parallel}}$ field sweep. The left hand inset in (b) shows sketches of PV density profiles across the crystal for increasing $H_{\parallel}$ (bottom to top). The right hand inset in (b) plots the vortex lensing amplitude  $(\Delta \widetilde{B}_z = \widetilde{B}_{z}^{max} (H_{\parallel}) - \widetilde{B}_{z}^{min} (H_{\parallel}))$ at $H_z$=5.3~Oe for the central sensor as a function of temperature.}
\label{fig1}
\end{figure}

{\it Experimental results.---\/} Our vortex lensing experiments
have been performed on an as-grown BSCCO superconducting
single crystal ($T_c = 91 $K, dimensions
~1mm$\times$0.75mm$\times$50$\mu$m). 
The changes in $B_z$ arising from PV lensing/antilensing were
detected using a $25 \mu m$ wirewidth micro-Hall probe array
patterned in a GaAs/AlGaAs 2D electron gas \cite{array}. The BSCCO
single crystal was positioned directly above our sensor and
secured with paraffin wax. The array has thirteen addressable
elements, of which twelve were situated at different positions
under the crystal and the remaining uncovered one acted as a
reference.  The sensor was driven by a $45 \mu A$ 32 Hz ac current
and the Hall voltages detected with a lock-in amplifier. The out-of-plane $H_z$ and in-plane
$H_{\parallel}$ magnetic field components are
varied independently using a solenoid and Helmholtz coil pair (one
of which could be precisely positioned vertically on a micrometer
driven stage), respectively. We use the in-plane `lock-in'
transition \cite{kosh93} to align the in-plane field within $\pm 
0.006^{\circ}$ of the \textit{ab} crystallographic planes.  The
exact alignment position is inferred from plots of
1/$H_{\parallel}$ at lock-in as a function of the height of the
movable Helmholtz coil, in a small fixed out-of-plane magnetic
field.
At the start of lensing experiments a fixed PV density was
established by field-cooling the BSCCO crystal from above $T_c$ in
a known value of the out-of-plane field, $H_z$. The in-plane
magnetic field $H_{{\parallel}}$ was then cycled several times
until a steady-state loop was obtained. During each cycle,
$H_{{\parallel}}$ was slowly (1.7~Oe s$^{-1}$) ramped up to a
maximum of 150~Oe and down to a minimum of -150~Oe and back to
zero, while the Hall voltage was monitored at a chosen element to
measure the magnetic induction, $B_z$.  In practice the observed
lensing was a function of the measurement position under the
crystal.  For the sake of brevity we only present data here for
two elements, one at the sample center and one $225\mu$m from one of the
edges parallel to $H_{\parallel}$, which fully illustrate the extremes of behaviour
seen. Fig. \ref{fig1} shows lensing
data measured (a) at the central location and (b) near the edge of
the sample at 77K for different values of $H_{z}$. In all cases
the data have been symmetrized $(\widetilde{B}_{z}\uparrow (H_{\parallel})=\frac{1}{2} 
(B_{z}\uparrow (H_{\parallel}) + B_{z}\downarrow
(-H_{\parallel})))$ to account for a small misalignment between
the plane of the BSCCO crystal and the Hall probe array.
In both cases for $H_z < 2$~Oe, the PV system shows a weak
reversible response which inverts when $H_z$ is inverted,
attributable to the dragging of PV stacks which are {\it all} trapped on JV chains.  For $H_z >$
2~Oe, free PVs exist between chains (mixed chains/lattice state) and
we start to see stronger, irreversible behaviour related to the compression of free PVs
and their cutting through JV stacks at high in-plane fields.
At higher $H_z$ the magnetization loop $\widetilde{B}_z(H_{\parallel})$
for the central element has a ``butterfly'' shape exhibiting: (i)
a fast increase of PV density when $H_{{\parallel}}$
increases from zero, followed by weaker (saturation-like)
dependence of $\widetilde{B}_z^{\rm central}(H_{{\parallel}})$; (ii) a rapid
reduction of $\widetilde{B}_z^{\rm central}(H_{{\parallel}})$ when
$H_{{\parallel}}$ decreases from its maximum value, followed by a
remarkable {\it antilensing} (an overshoot in the reduction of PV
density) effect $\widetilde{B}_z^{\rm central}(H_{{\parallel}}>0)<\widetilde{B}_z^{\rm
central}(H_{{\parallel}}=0)$. Note that on the
$H_{{\parallel}}$-increasing branch of the lensing loop, $\widetilde{B}_z^{\rm
central}(H_{{\parallel}})$ is always higher than on the decreasing
branch on the right-side ($H_{{\parallel}}>0$) of the hysteresis cycle, i.e., $\widetilde{B}_z^{\rm
central}(H_{{\parallel}},\,dH_{{\parallel}}/dt>0)>\widetilde{B}_z^{\rm
central}(H_{{\parallel}},\,dH_{{\parallel}}/dt<0)$. We
denote such loops as ``clockwise''.

The data from the Hall element near the sample edge (Fig.
\ref{fig1}b) provide insights into the spatial distribution of
the PV density in our lensing experiments. In stark contrast to
Fig. \ref{fig1}a, we now see strong antilensing behaviour for
$H_z<6.5$\,Oe. This is easily understood in terms of the PV
profiles generated during our experiments. Since PVs are pushed
from two opposite edges of the sample towards the center, there
must be regions near these edges which experience a decrease in PV
density, while accumulation is occurring in the crystal center
(see sketched profiles in the left hand inset of Fig. \ref{fig1}b).
It is interesting to note that the counter-clockwise loops at low
$H_z$ transform to clockwise ones when $H_z$ increases. This is
best illustrated in the curve at $H_z = 6.5$ \,Oe, which shows an extra 
crossing point for $H_{\parallel}>0$ between traces on the
sweep-up and sweep-down.
At such high values of $H_z$ PVs penetrating through the sample surface partially
compensate for the deficit of PVs near the edges, and enhanced PV-PV repulsion results in a broadening of the
focus region and a shift of the lensing/antilensing interface
towards the sample edge.  Hence a transition from
counter-clockwise to clockwise $\widetilde{B}_z$ loops occurs as the
out-of-plane field is increased.  We conclude that the PV profiles,
and their changes, are a subtle function of the applied
out-of-plane field, allowing us to precisely control the ``magnetic
landscape'' inside the sample. Vortex control is achieved here
using just the so-called dc-driven (quasi-adiabatic) mode;
ac-drives with trains of time-asymmetric $H_{{\parallel}}$ pulses
will be described elsewhere \cite{natmat}.

The right hand insert of Fig. \ref{fig1}b shows a plot of the lensing amplitude 
at $H_z$=5.3~Oe as a function of temperature.  Surprisingly, we find that the lensing amplitude
falls rapidly with temperature, being very weak at 84~K and
undetectable at 86~K.  Since this is well below the critical
temperature of the BSCCO crystal ($T_c \sim$ 91K), and the crossing
lattices interaction strength is only very weakly temperature
dependent \cite{kosh}, we conclude that efficient lensing requires {\it finite} bulk pinning
to prevent PVs escaping from focus regions laterally
parallel to JV stacks. The effectiveness of bulk pinning will drop rapidly at
elevated temperatures leading to a rapid reduction in efficiency.

\begin{figure}
\vspace*{-0.5cm}
\includegraphics[width=0.96 \linewidth]{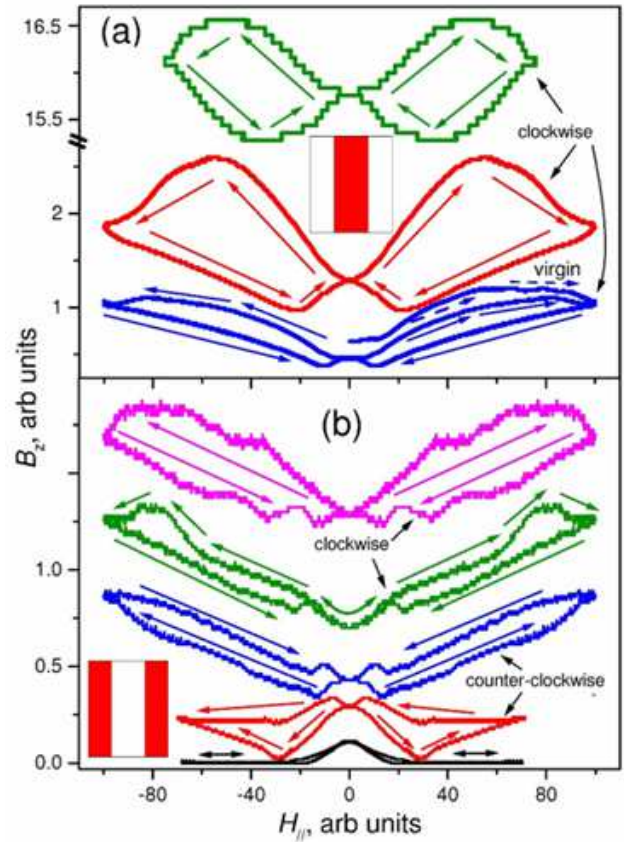}
\caption{(Color online) Simulated loops of the local induction
$B_z(H_{{\parallel}})$ at the center (a) and near the sample edges
(b) for different values of $H_z$ ($H_z$ increasing from 
bottom to top in each panel). The red region in each inset
indicates where $B_z$ was monitored in the sample. The main
features found in experiments (Figs. \ref{fig1}a, \ref{fig1}b) are
reproduced in the simulations.}\label{fig3}
\end{figure}

{\it Simulations.---} The minimal model to simulate the observed
lensing effects describes the overdamped dynamics of JV and PV
rows using a set of coupled equations of motion: $\gamma\,\eta_J\,
\dot{x}^J_i/a^J=f^{JJ}_i+f^{JH}_i+f^{JP}_i$, and $\eta_P\,
\dot{x}^{P}_k/a^P=f^{PP}_k+ f^{PH}_k+f^{PJ}_k$, where $x^{J}_i$
and $x^{P}_{k}$ are the positions of JV and PV rows with distances
between JVs(PVs) in a row $a^J/\gamma$($a^P$) and
anisotropy parameter $\gamma$, while $\eta_J$ and $\eta_{P}$ are
the JV and PV viscosities. The viscous forces slowing down the
vortex motion are balanced by: (1) the repulsive force $f^{JJ}$
between JV rows (including images of rows with respect to the
sample surface); (2) the interaction $f^{JH}$ of JV rows with
Meissner currents generated by the externally-applied
time-dependent magnetic field $H_{{\parallel}}(t)$; (3) the
repulsion $f^{PP}$ between PV rows (including images); (4) the
interaction $f^{PH}$ of PV rows with $H_z$; and (5) the attractive
forces $f^{JP}$ between rows of JVs and PVs.

The interaction between vortex rows decays
exponentially~\cite{sav-micro}:
$f^{JJ}_i\,\tau/D\,\eta_J=(a^P\beta/a^J)\sum_j\sgn[x^J_i-x^J_{j}]\exp[-|x^J_i-x^J_{j}|/\lambda_c]$,
$f^{PP}_k\tau/D\,\eta_P=\sum_j\sgn[x^P_k-x^P_{j}]\exp[-|x^P_k-x^P_{j}|/\lambda_{ab}]$,
where the interaction lengths are the in-plane $\lambda_{ab}$ and
$c$-axis $\lambda_c$ penetration lengths, while
$\beta=\eta_P/(\gamma\eta_J)$ is the relative viscosity.
Hereafter, we normalize all distances by the half-width of the
sample $D$ and all time scales by
$\tau=16\pi\lambda_{ab}^2a^P\eta_PD/\Phi_0^2$. The distances
between PV and JV rows are related to the $H_z$ and
$H_{{\parallel}}$ magnetic field components by
$a^P\approx({\Phi_0/H_z})^{1/2}$ and $a^J\approx
({\gamma\Phi_0/H_{{\parallel}}})^{1/2}$. The interaction with the
Meissner current decays exponentially on the scales $\lambda_c$
and $\lambda_{ab}$ from the surface ($x=\pm 1$) of the sample, for
JVs and PVs, respectively. The corresponding forces can be
modelled \cite{sav-micro} as $f^{JH}_i\tau/D\eta_J = -
(2\lambda_ca^P\beta/(a^J)^2) \sinh(x^J_i/\lambda_c) /
\cosh(D/\lambda_c)$ and $f^{PH}_k\tau/D\,\eta_P = -
(2\lambda_{ab}/a^P) \sinh(x^P_k/\lambda_{ab}) /
\cosh(D/\lambda_{ab})$. The JV-PV attraction can be approximated
as $f^{JP}_i\tau/D\,\eta_J =
\lambda_{ab}^2\beta\beta_1/(a^J)^2\sum_k\sgn(x^P_k-x^J_i)
\exp(-4\pi|x^P_k-x^J_i|/a^J)$ and
$f^{PJ}\tau/D\,\eta_P\,=(\lambda_{ab}^2a^P\beta_1/(a^J)^3)\sum_i\sgn(x^J_i-x^P_k)
\exp(-4\pi|x^P_k-x^J_i|/a^J)$, where $\beta_1\approx
16\pi^2/\ln\bigl\{1+[(\lambda_c^2/a^J)^{2}+1]/[(\lambda_{ab}/a^P)^{2}+1]\bigr\}$
is related to the tilt elasticity of PV stacks.  The qualitative
agreement between the very complex and nontrivial experimental
data and simulations indicates that the model used captures the
essential physics.

Our molecular dynamics simulations follow the experiments. First
$H_{{\parallel}}$ was slowly increased from zero to
$H_{{\parallel}}^{\max}$, and then decreased back to zero over the
same period of time. Such cycles were repeated several times to
achieve steady-state loops. The average PV density was monitored
(as a function of $H_{{\parallel}}$) both at the center and near
the edges of the sample.

\begin{figure}
  \includegraphics[width=0.96 \linewidth]{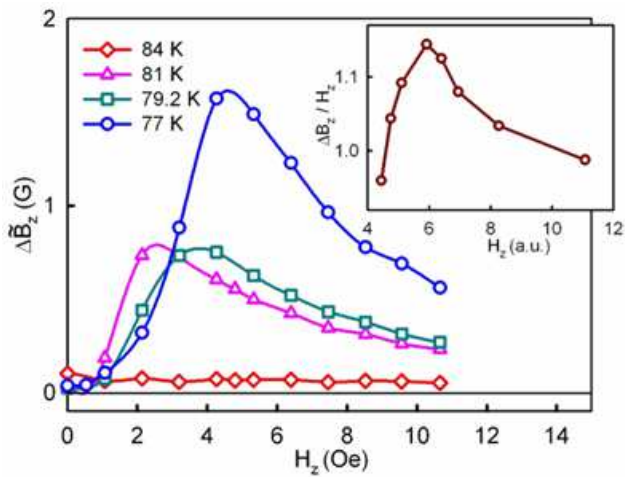}
  \caption{Net change of magnetic induction, $\Delta \widetilde{B}_z$, 
as a function of $H_z$ measured at the sample center for four
different temperatures.  Inset shows the simulated lensing efficiency.  Both experiment and simulation exhibit a
maximum as a function of $H_z$.}\label{fig4}
\end{figure}

{\it Comparison theory-experiment.---} The simulated clockwise
$B_z(H_{{\parallel}})$ ``butterfly'' loops for the PV
density at the center of the sample show the same features
\cite{anim} that were observed in experiments (c.f. 
Figs. \ref{fig1}a and \ref{fig3}a) and can be easily interpreted: (i) First, $\widetilde{B}_z^{\rm central}$
increases with $H_{{\parallel}}$ as JVs move towards the sample center and drag PVs with them.
This is consistent with theoretical
predictions \cite{nature_materials}; (ii) At a certain in-plane
field $H_{{\parallel}}$, the PV density at the center of the
sample saturates and even starts to decrease.  The PV density at the center is now
large enough that PV-PV repulsion becomes very large. PVs start to cut through the JV rows
and, near the maximum compression point, some of these PVs escape the narrow potential produced by the JVs and are lost;
(iii) On the decreasing branch of the loop, both experiments and simulations exhibit a remarkable
anti-lensing effect.  This arises because a smaller total number
of PVs now spreads out over the whole sample, resulting in a deficit
of PVs at the center. Also the ratio of the lensing to antilensing
effect, $\max[\widetilde{B}_z^{\rm central}(H_{{\parallel}})-\widetilde{B}_z^{\rm
central}(0)]/|\min[\widetilde{B}_z^{\rm central}(H_{{\parallel}})-\widetilde{B}_z^{\rm
central}(0)]|$ decreases to about one when the out-of-plane field
increases, in agreement with experiments (this produces rounder
loops for higher $H_z$, see Figs. \ref{fig1}a,
\ref{fig3}a).

The simulations also capture all the main features found in the
experiments near the sample edge: (i) At very low out-of-plane
fields, $H_z$, the area of this remarkable counterclockwise loop
increases with $H_{z}$; (ii) at higher out-of-plane fields the
counterclockwise loops narrow, transform into clockwise loops and
broaden again.

Fig. \ref{fig4} illustrates the experimental (main figure) and
simulated (inset) dependence of lensing amplitude on
out-of-plane magnetic field.  In both cases the amplitude shows a
pronounced peak as a function of $H_z$.  At 77K the experimental
lensing efficiency exhibits a maximum of nearly 40\% at
$H_z\sim5$\,Oe. This field value is in reasonable agreement with the
predicted maximum in JV pinning strength (by PVs) at $B_{z}\sim
0.26 \frac{\Phi_{0}}{(\gamma s)^{2}}\sim 6$G \cite{koshelev2003}.

{\it Conclusions.---} We have experimentally realized concave and
convex magnetic vortex lenses which focus and defocus the
out-of-plane magnetic flux in a \bis sample. Remarkably, this was
done by employing the ``dragging'' of PVs by JVs with {\it no fixed
spatial asymmetry} in the sample. We show that the PV density near
the center/edges of the sample can easily be controlled by
changing either the in-plane or the out-of-plane fields, as well
as by varying the temperature. The experimental results are well
described by a simple model considering the dragging of one vortex
species by the other. This novel method of
quantum-motion-control ({\it a ratchet with no ratchet potential}) opens
up new avenues for the manipulation of flux quanta and nanoscale
particles.

We acknowledge support from EPSRC in the UK under grant No. GR/R46489/01,
the ESF VORTEX network, the US NSA and ARDA under AFOSR contract No. F49620-02-1-0334,
NSF grant No. EIA-0130383, and a Grant-in-Aid from MEXT, Japan.


\begin{references}

\bibitem{baert95} M. Baert {\it et al.,} Phys. Rev. Lett. {\bf74}, 3269
(1995).

\bibitem{martin97} J.I. Martin {\it et al.,} Phys. Rev. Lett. {\bf79}, 1929
(1997).

\bibitem{morgan98} D.J. Morgan {\it et al.,} Phys. Rev. Lett. {\bf80}, 3614
(1998).

\bibitem{bael99} M.J. Van Bael {\it et al.,} Phys. Rev. B {\bf59}, 14674
(1999).

\bibitem{lyuksyutov98} I.F. Lyuksyutov {\it et al.,} Phys. Rev. Lett. {\bf81}, 2344 (1998).

\bibitem{vortex} J.F. Wambaugh {\it et al.,} Phys. Rev. Lett. {\bf 83}, 5106 (1999); 
C. J. Olson {\it et al.,} ibid {\bf 87}, 177002 (2001); 
B.Y. Zhu {\it et al.,} ibid {\bf 92}, 180602 (2004).

\bibitem{kwok-2003}  W.K.~Kwok, {\it et al.,} Physica C {\bf 382}, 137 (2002).

\bibitem{ourscience} J.E. Villegas {\it et al.,} Science {\bf 302}, 1188 (2003).

\bibitem{moshchalkov} J. Van de Vondel {\it et al.,} Phys. Rev. Lett. {\bf 94}, 057003 (2005).

\bibitem{togawa} Y. Togawa {\it et al.,}  Phys. Rev. Lett. {\bf 95}, 087002 (2005).

\bibitem{roger} R. W\"ordenweber {\it et al.,} Phys. Rev. B {\bf 69}, 184504 (2004).

\bibitem{nature_materials} S. Savel'ev and F. Nori, Nature Materials {\bf 1}, 179 (2002).

\bibitem{prl} S. Savel'ev {\it et al.,} Phys. Rev. Lett. {\bf 92}, 160602 (2004).

\bibitem{tonomura} T. Matsuda, {\it et al.,}  Science {\bf 294}, 2136 (2001).

\bibitem{bending} A. Grigorenko {\it et al.,} Nature {\bf 414}, 728 (2001).

\bibitem{vlasko} V.K. Vlasko-Vlasov {\it et al.,} Phys. Rev. B {\bf 66}, 014523 (2002).

\bibitem{tamegai} M. Tokunaga {\it et al.,} Phys. Rev. B {\bf 66}, 060507(R) (2002).

\bibitem{bolle91} C.A. Bolle et al., Phys. Rev. Lett. {\bf66}, 112-115 (1991).

\bibitem{bul} L.N. Bulaevskii {\it et al.,} Phys. Rev. B {\bf 46}, 366 (1992).

\bibitem{kosh} A.E. Koshelev, Phys. Rev. Lett. {\bf 83}, 187 (1999).

\bibitem{sav} S.E. Savel'ev {\it et al.,} Phys. Rev. B {\bf 64}, 094521 (2001).

\bibitem{icl} G.K. Perkins {\it et al.,} Supercond. Sci. Technol. {\bf18}, 1290 (2005)

\bibitem{array} M.S. James {\it et al.,} Phys. Rev. B {\bf 56}, R5771 (1997).

\bibitem{kosh93} A.E. Koshelev, Phys. Rev. B {\bf 48}, 1180 (1993).

\bibitem{natmat} D. Cole {\it et al.,} Nature Materials (in press).

\bibitem{sav-micro} S.E. Savel'ev, V.S. Gorbachev {\it et al.,} JETP {\bf 83}, 570 (1996).

\bibitem{anim} Animations illustrating vortex lensing are 
available  at http://dml.riken.go.jp/vortex-dc

\bibitem{koshelev2003} A.E. Koshelev, Phys. Rev. B {\bf68}, 094520 (2003).

\end{references}
\end{document}